\documentclass[prb,twocolumn,showpacs,amsmath,amssymb,superscriptaddress]{revtex4-2}
\usepackage{dcolumn}
\usepackage{bm,graphicx}
\usepackage{etoolbox}
\usepackage{url}
\usepackage{booktabs}
\usepackage[colorlinks]{hyperref}
\usepackage{breakurl}
\usepackage{color}
\usepackage [latin1]{inputenc}
\usepackage{siunitx}

\newsavebox\myboxA
\newsavebox\myboxB
\newlength\mylenA

\newcommand*\xoverline[2][0.75]{%
    \sbox{\myboxA}{$\m@th#2$}%
    \setbox\myboxB\null
    \ht\myboxB=\ht\myboxA%
    \dp\myboxB=\dp\myboxA%
    \wd\myboxB=#1\wd\myboxA
    \sbox\myboxB{$\m@th\overline{\copy\myboxB}$}
    \setlength\mylenA{\the\wd\myboxA}
    \addtolength\mylenA{-\the\wd\myboxB}%
    \ifdim\wd\myboxB<\wd\myboxA%
       \rlap{\hskip 0.5\mylenA\usebox\myboxB}{\usebox\myboxA}%
    \else
        \hskip -0.5\mylenA\rlap{\usebox\myboxA}{\hskip 0.5\mylenA\usebox\myboxB}%
    \fi}
\makeatother
\usepackage[export]{adjustbox}
\usepackage{float}
\usepackage{siunitx}
\usepackage{booktabs}
\usepackage{color}
\usepackage{xcolor}
\usepackage [latin1]{inputenc}
\usepackage{dcolumn}
\usepackage{etoolbox}
\usepackage{url}
\usepackage[colorlinks]{hyperref}
\usepackage[normalem]{ulem}
\usepackage{array}
\newcolumntype{x}[1]{>{\centering\arraybackslash\hspace{0pt}}p{#1}}
\usepackage{multirow}
\usepackage{accents}
\usepackage{braket}

\usepackage[normalem]{ulem}

\begin{document}


\title{Dynamics of surface electrons in a topological insulator:\\ cyclotron resonance  at room temperature}

\author{I.~Mohelsky}
\email{ivan.mohelsky@centrum.cz}
\affiliation{Laboratoire National des Champs Magn\'etiques Intenses, EMFL, CNRS UPR3228, Univ.~Grenoble Alpes, Univ.~Toulouse, Univ.~Toulouse~3, INSA-T, Grenoble, France}
\affiliation{Department of Physics, Universit\'e de Fribourg, Chemin du Mus\'ee 3, Fribourg, 1700 Switzerland}

\author{F.~Le Mardel\'e}
\affiliation{Laboratoire National des Champs Magn\'etiques Intenses, EMFL, CNRS UPR3228, Univ.~Grenoble Alpes, Univ.~Toulouse, Univ.~Toulouse~3, INSA-T, Grenoble, France}

\author{J.~Dzian}
\affiliation{Laboratoire National des Champs Magn\'etiques Intenses, EMFL, CNRS UPR3228, Univ.~Grenoble Alpes, Univ.~Toulouse, Univ.~Toulouse~3, INSA-T, Grenoble, France}
\affiliation{Faculty of Mathematics and Physics, Charles University, Ke Karlovu 5, Prague, 121 16, Czech Republic}

\author{J.~Wyzula}
\affiliation{Laboratoire National des Champs Magn\'etiques Intenses, EMFL, CNRS UPR3228, Univ.~Grenoble Alpes, Univ.~Toulouse, Univ.~Toulouse~3, INSA-T, Grenoble, France}
\affiliation{Scientific Computing, Theory and Data Division, Paul Scherrer Institute, Switzerland}

\author{X.~D.~Sun}
\affiliation{Laboratoire National des Champs Magn\'etiques Intenses, EMFL, CNRS UPR3228, Univ.~Grenoble Alpes, Univ.~Toulouse, Univ.~Toulouse~3, INSA-T, Grenoble, France}

\author{C.~W.~Cho}
\affiliation{Laboratoire National des Champs Magn\'etiques Intenses, EMFL, CNRS UPR3228, Univ.~Grenoble Alpes, Univ.~Toulouse, Univ.~Toulouse~3, INSA-T, Grenoble, France}

\author{B.~A.~Piot}
\affiliation{Laboratoire National des Champs Magn\'etiques Intenses, EMFL, CNRS UPR3228, Univ.~Grenoble Alpes, Univ.~Toulouse, Univ.~Toulouse~3, INSA-T, Grenoble, France}

\author{M.~Shankar}
\affiliation{Institute of Physics, Academia Sinica, Nankang, Taipei, 11529, Taiwan}

\author{R.~Sankar}
\affiliation{Institute of Physics, Academia Sinica, Nankang, Taipei, 11529, Taiwan}

\author{A.~Ferguson}
\affiliation{Department of Physics, Universit\'e de Fribourg, Chemin du Mus\'ee 3, Fribourg, 1700 Switzerland}

\author{D.~Santos-Cottin}
\affiliation{Department of Physics, Universit\'e de Fribourg, Chemin du Mus\'ee 3, Fribourg, 1700 
Switzerland}

\author{P.~Marsik}
\affiliation{Department of Physics, Universit\'e de Fribourg, Chemin du Mus\'ee 3, Fribourg, 1700 Switzerland}

\author{C.~Bernhard}
\affiliation{Department of Physics, Universit\'e de Fribourg, Chemin du Mus\'ee 3, Fribourg, 1700 Switzerland}

\author{A.~Akrap}
\affiliation{Department of Physics, Universit\'e de Fribourg, Chemin du Mus\'ee 3, Fribourg, 1700 Switzerland}
\affiliation{Department of Physics, Faculty of Science, University of Zagreb, 10000 Zagreb, Croatia}

\author{M.~Potemski}
\affiliation{Laboratoire National des Champs Magn\'etiques Intenses, EMFL, CNRS UPR3228, Univ.~Grenoble Alpes, Univ.~Toulouse, Univ.~Toulouse~3, INSA-T, Grenoble, France}
\affiliation{Institute of High Pressure Physics, PAS, Warsaw, PL-01-142 Poland}
\affiliation{CENTERA, CEZAMAT, Warsaw University of Technology, Warsaw, PL-02-822 Poland}

\author{M.~Orlita}
\email{milan.orlita@lncmi.cnrs.fr}
\affiliation{Laboratoire National des Champs Magn\'etiques Intenses, EMFL, CNRS UPR3228, Univ.~Grenoble Alpes, Univ.~Toulouse, Univ.~Toulouse~3, INSA-T, Grenoble, France}
\affiliation{Faculty of Mathematics and Physics, Charles University, Ke Karlovu 5, Prague, 121 16, Czech Republic}
    
\date{\today}

\begin{abstract} 
The ability to manipulate the surface states of topological insulators using electric or magnetic fields under ambient conditions is a key step toward their integration into future electronic and optoelectronic devices. Here, we demonstrate -- using cyclotron resonance measurements on a tin-doped BiSbTe$_2$S topological insulator -- that moderate magnetic fields can quantize massless surface electrons into Landau levels even at room temperature. This finding suggests that surface-state electrons can behave as long-lived quasiparticles at unexpectedly high temperatures.
\end{abstract}

\maketitle
Exploring the non-trivial topology of electronic states in solids is nowadays pursued as a path to discover new quantum phases of matter, enabling robust electronic and spintronic applications, advancing quantum computing, or uncovering fundamental connections between condensed matter and high-energy physics~\cite{HasanRMP10,QiRMP11,ArmitageRMP18}. Despite initial expectations that topological protection might confer a certain robustness against increasing temperature, this research often remains limited to cryogenic temperatures. While conventional thermally induced dissipation processes may affect topological materials in much the same way as other solids, the stability of topological phases at elevated temperatures is far from trivial, as frequently discussed in theoretical studies~\cite{RivasPRB13,GaratePRL13,Viyuela2DM15,KempkesSR16, MoligniniPRR23}.  

Experimentally, the ARPES technique applied to topological insulators (TIs) is likely the only technique that provided clear evidence for the existence of topological surface states under ambient conditions~\cite{KumarPRB22}. Optical spectroscopy techniques, central to this work, along with methods such as electronic transport, are directly relevant to the applications of topological materials. Although these techniques have been frequently used to probe surface states~\cite{SchafgansPRB12,AssafSR16,PhuphachongC17}, this has primarily been limited to low temperatures. Demonstrating the existence of surface states under ambient conditions using these methods is therefore crucial for advancing electronic and optoelectronic devices based on topological materials.

\begin{figure}[b]
      \includegraphics[width=.45\textwidth]{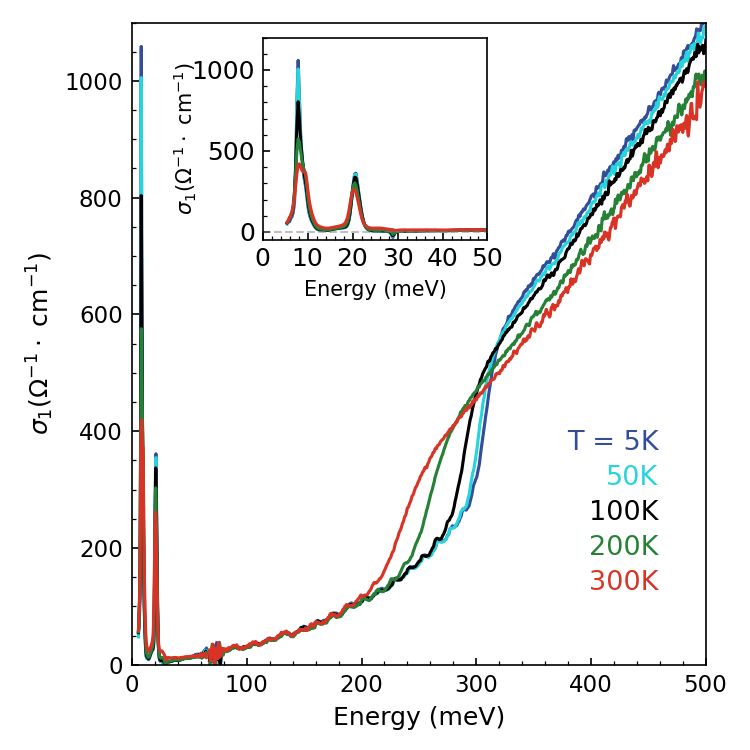}
      \caption{\label{Ellipsometry} The real part of optical conductivity of Sn-BST2S deduced by ellipsometry at $T= 5, 50, 100, 200$ and 300~K. The oscillations below the band gap correspond to a delaminated layer with the thickness of $(7 \pm 2)$~$\mu$m. The inset shows the low-energy part of the spectrum with two pronounced infrared-active phonons, but no apparent Drude contribution.}
\end{figure}

In this study, we target BiSbTe$_2$S (BST2S), a topological insulator in which tin-induced in-gap states ensure its bulk-insulating nature~\cite{KushwahaSR16,GudacCM24} while preserving a high electronic quality. This is illustrated, e.g., by electron transport through surface states in the regime of the quantum Hall effect (QHE), observed at liquid helium temperatures~\cite{IchimuraAPL19,MatsushitaPRM21}. Here we identify bulk BST2S crystals as a useful testbed for optical and magneto-optical studies of topologically protected surface states in TIs. 
We show, experimentally, that the surface states of a TI can be at room temperature controlled by a moderate magnetic field and even the regime of full Landau quantization is achievable. The electron-phonon interaction, otherwise effective for bulk states in narrow-gap semiconductors such as BiSbTe$_2$S, appears to affect weakly the surface electrons.

\begin{figure*}
      \includegraphics[width=0.99\textwidth]{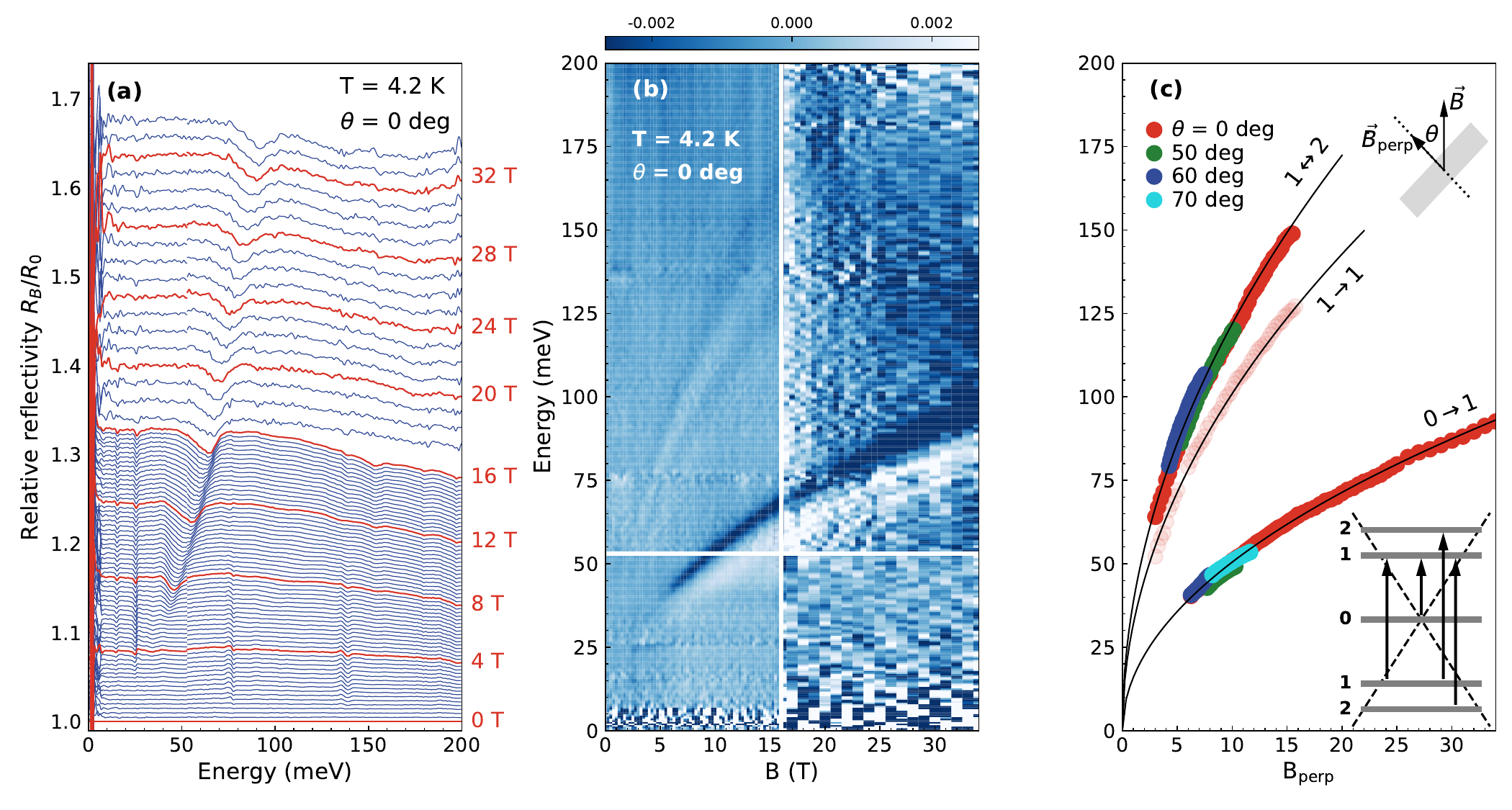}
      \caption{\label{Magneto-optics} Low-temperature infrared magneto-reflectivity of BST2S: (a) Stack-plot of magneto-reflectivity spectra, $R_B/R_0$, and (b) false-color plot of the corresponding derivatives, $d/dB [R_B/R_0]$. (c) The energies of observed inter-LL excitations as a function of the magnetic-field component perpendicular to the sample ($B_{\mathrm{perp}}$), extracted as the minima in the $R_B/R_0$ spectra collected in the tilted-field configuration (panel c, upper inset) for the angles $\theta = 0, 50, 60$ and 70~deg. The error bars are given by the size of symbols. 
      Black lines are the fits of the inter-LL transitions, using the model of massless Landau-quantized electrons, yielding the velocity parameter $v=(4.4\pm0.2)\times10^5$~m/s. The observed inter-LL excitations are indicated in the lower inset of panel c.}
\end{figure*}

To prepare Sn-doped BST2S crystals, the corresponding mixture of high-purity (5N) raw materials were grounded and then sealed in a quartz tube under the vacuum of about \SI{3e-2}{}~Torr. The initial powder was calcinated at \SI{990}{\kelvin} for \SI{18}{\hour}, reground and then heated at \SI{1100}{\kelvin} for \SI{12}{\hour} in a vertical Bridgman furnace to grow single crystals. The final atomic constitution was determined  using field-emission electron-probe microanalyzer (EPMA) as Bi$_{1.2}$Sb$_{0.7}$Te$_{2}$S with
the relative atomic composition of tin reaching 0.006 and its spatial variation up to 20\%. This translates into 8-12 tin atom for every 2000 atoms of bismuth. The magneto-transport characterization of one selected BST2S crystal is presented in Supplemental material~\cite{SM}.

\phantom{\cite{StephenSR20,HikamiPTP80,Gracia-AbadNM21}}

The synthetized BST2S crystals were characterized using the ellipsometry technique at selected temperatures (Fig.~\ref{Ellipsometry}). The observed response is typical of a narrow-gap semiconductor, with an onset of interband absorption (the band gap) around 300~meV at low temperatures. A pronounced tail of sub-gap excitations is also present. At low energies, two pronounced phonon lines are observed at 8 and 20~meV. They are relatively broad, as expected for phonons in a mixed compound. Based on the symmetry analysis~\cite{Richterpssb77,IvantchevJAC00}, we assign them to the $E_u$ infrared-active modes. Importantly, we find no signs of free-charge-carrier (Drude-type) absorption. This is in line with findings of the preceding study by Jiang et al.~\cite{JiangPRB20}, except for the maximum in the optical conductivity at the absorption edge. This feature was interpreted in terms of a van-Hove singularity in the joint density of states due to Mexican-hat-type profile of bands at the $\Gamma$ point, but it is completely absent in our ellipsometry data. 

\begin{figure*}
      \includegraphics[width=0.99\textwidth]{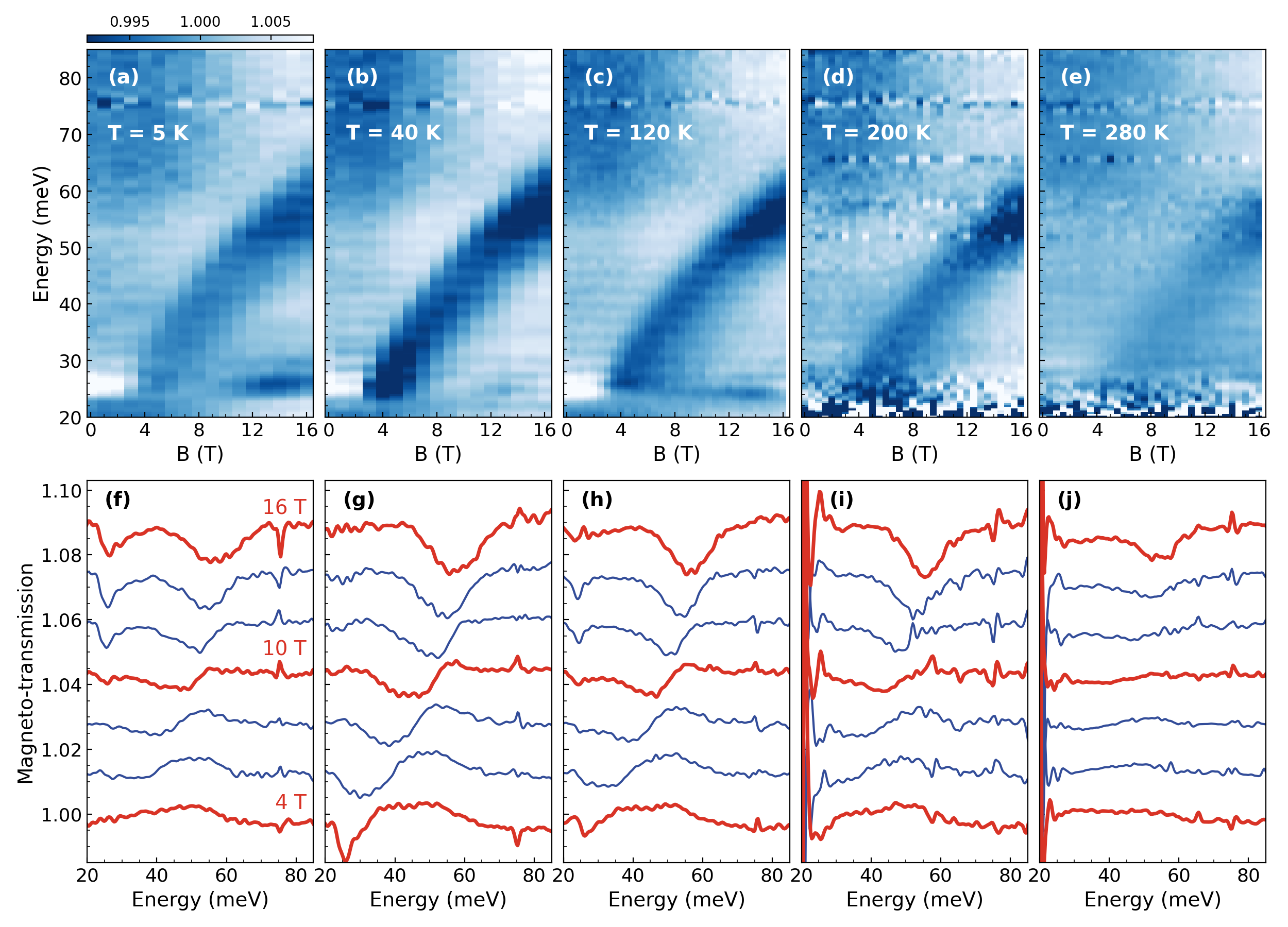}
      \caption{\label{temp} Temperature evolution of the fundamental CR line: (a-e) Relative (double) magneto-transmission of a 7-$\mu$m-thick BST2S layer, $T_B/T_0$, for selected temperatures plotted in a form of false-color plots. (f-j) Magneto-transmission spectra at selected temperatures and magnetic fields applied.} 
\end{figure*}

Then, a series of magneto-optical optical experiments was performed on bulk crystals. In all presented measurements, the reflectivity was measured in the Faraday or tilted-field configurations. The radiation of a globar was analyzed by the Bruker Vertex 80v Fourier-transform spectrometer and delivered via light-pipe optics to the sample located in a superconducting and resistive coil, below and above 16~T, respectively. A part of the reflected signal was deviated, using a silicon beamsplitter, toward an external bolometer. A set of relative magneto-reflectivity spectra, $R_B/R_{B=0}$, collected on a bulk BST2S sample in the Faraday configuration, with $B$ oriented along the $c$ axis, is presented in Fig.~\ref{Magneto-optics}a. For these measurements, we used a plate-like crystal with the approximate thickness of 1~mm and probed the area of 3$\times$3~mm$^2$. Before a series of experiments, the crystal was stored under ambient conditions and its surface was not protected or treated apart from the initial exfoliation. 

The observed response is dominated by a single excitation that emerges above $B\approx6$~T at the energy of 40~meV, \emph{i.e.}, above the phonon modes. It monotonically blueshifts up to the highest applied magnetic field (34~T). The $B$-derivative of the data, $d[R_B/R_0]/dB$, presented as a false color plot in Fig.~\ref{Magneto-optics}b, reveals additional two lines at higher energies. These are clearly visible in low magnetic fields, but cannot be resolved in the data collected using the resistive coil because of a lower signal-to-noise ratio. Notably, all excitations are observed  at energies far below the bulk band gap of BST2S.

The interpretation of our data is greatly facilitated by the knowledge accumulated in preceding magneto-optical studies of electrons in conical bands. These are found, e.g., in graphene, but also in many other materials~\cite{SadowskiPRL06,OrlitaPRL08II,GusyninPRL07, JiangPRL07,OrlitaSST10,ChenPRL15,LuSA22,Santos-CottinPRB22}. The energy spacing of the observed transitions, $1:2:1+\sqrt{2}$, as well as their characteristic $\sqrt{B}$-dependence allow us to assign them to particular excitations within the Landau level (LL) spectrum of Dirac electrons, $E_n^{\pm}=\pm \sqrt{(\Delta/2)^2+2v^2e\hbar n B}$, $n=0,1,2\ldots$, in which the asymptotic velocity reaches $v=(4.4 \pm 0.2)\times10^5$~m/s and the gap parameter $\Delta$ vanishes (with the upper limit of several meV). The most pronounced line in our data corresponds to the fundamental CR mode, $0(1)\rightarrow 1(0)$, see the lower inset of Fig.~\ref{Magneto-optics}c. 

The conical bands are typical of many topological materials. Notably, they represent the most salient part of bulk electronic band structure of three-dimensional Dirac and Weyl semimetals~\cite{ArmitageRMP18}.  In contrast, in topological insulators -- which behave as conventional narrow-gap semiconductors in their bulk -- conical bands can only emerge at the surface. This allows us to associate the observed response with the topological protected gapless surface states in BST2S. The magneto-reflectivity experiments performed in the tilted-field configuration, see the upper inset of Fig.~\ref{Magneto-optics}c, are in line with this assignment, thus illustrating a response sensitive only to the magnetic-field component perpendicular to the sample surface. Importantly, the vanishing parameter $\Delta$ excludes that the observed inter-LL resonances might stem from massive (gapped) Dirac states on the surface, referred to as Volkov-Pankratov states~\cite{VolkovJETP85,LuEPL19}. Such a low value of $\Delta$ might only emerge at a very smooth interface between a topological and normal insulator~\cite{LuEPL19}, not at the atomically sharp surface.

The characteristic pattern due to the Landau-quantized massless electrons on the surface of BST2S was observed in reflectivity, but also in transmission discussed later on. In particular, the most pronounced $0(1)\rightarrow 1(0)$ excitation appeared in a series of explored samples, with the same characteristic $\sqrt{B}$ dependence, and within the experimental linewidth, also at the same position in the spectra. However, the integral intensity varied. For certain specimens, the response due to surface electrons was not resolved at all. We interpret this in terms of an occupation effect. The distance of the Fermi energy from the Dirac point is not a directly controlled parameter in our experiments. Therefore, the carrier density in the surface states may vary among different samples and even at different locations of one sample. We expect that it is defined by the properties of underlying bulk (band bending effects, in-gap states, charged impurities), as well as by the interaction of the crystal surface with environment (e.g., adsorbed atom/molecules). Similar to graphene~\cite{MartinNP08}, electron-hole puddles, or in a broader sense, the spatial variation of the charge density in the surface states of TIs are regularly observed~\cite{BeidenkopfNP11,SkinnerPRB13}. 

Preceding transport studies~\cite{KushwahaSR16,IchimuraAPL19} indicated that the surface states exhibit an electron-type conductivity. The appearance of the electron-like $0 \rightarrow 1$ resonance around $B\approx6$~T (Fig.~\ref{Magneto-optics}) indicates that $n=1$ LL is no longer fully occupied, i.e., the filling factor dropped below $\nu=3/2$ at least in certain locations. There, the electron density is as low as $\nu\cdot\zeta\approx 2\times10^{11}$~cm$^{-2}$ where $\zeta=eB/h$ is the LL degeneracy. With a further increase of $B$ -- when the filling factor decreases below 1/2 -- also the hole-like $1 \rightarrow 0$ transition becomes observable (cf. inset in Fig.~\ref{Magneto-optics}).

The lines at higher energies are interband inter-LL excitations: namely, the twice degenerate $1(2)\rightarrow 2(1)$ transition accompanied by the $1\rightarrow 1$ line that is symmetric across the $n=0$ LL. Interestingly, this latter line is not expected among the transitions following the basic set of selection rules, $n \rightarrow n \pm 1$, for electric-dipole-active inter-LL excitations. Nevertheless, it has been observed in the response of high-quality Landau-quantized graphene and interpreted as due to point defects, strain, random Coulomb potential, electron-electron interactions or a combination of these factors~\cite{NedoliukNT19}. The observed response shows no indication of any electron-hole asymmetry or a deviation of the conical band from linearity. The former would lift the double degeneracy of the $n\rightarrow n+1$ and $n+1\rightarrow n$ transitions ($n=0$ and 1, in our case) and the latter would imply a departure from the $\sqrt{B}$ dependence, respectively, see Ref.~\cite{PlochockaPRL08}.

Let us now turn our attention to another set of magneto-optical data presented in Fig.~\ref{temp}. These were collected at temperatures ranging from 5 to 280~K, using the experimental arrangement described in Ref.~\cite{MohelskyPRB23}. To maximize the signal in the reflectivity configuration, a thin slab (thickness $\approx7$~$\mu$m) was deposited on a gold mirror and the collected data thus may be viewed as double-pass magneto-transmission spectra. The data collected at selected temperatures are presented in a form false-color plots (Figs.~\ref{temp}a-e). Each color plot is accompanied by a set of magneto-transmission spectra measured at several values of the applied magnetic field (Figs.~\ref{temp}f-j).  

The magneto-optical response traced in Fig.~\ref{temp} is dominated by a single magneto-transmission resonance. 
Its characteristic $\sqrt{B}$ dependence (see the further analysis in~\cite{SM}), as well as its position -- within the linewidth the same as in Fig.~\ref{Magneto-optics} -- allows us to identify it as the $0(1) \rightarrow 1(0)$ transition. Hence, the fundamental CR mode of Landau-quantized electrons on the surface of BST2S can be at moderate magnetic fields observed up to room temperature. A small redshift of the CR line is observed with increasing $T$. This indicates a small decrease, approximately by 5\%,  of the velocity parameter, as compared to its low-temperature value (see also the analysis in~\cite{SM}).  

The width of the transition (FWHM $\approx$ 15~meV) is smaller that its position in the spectrum. The visibility of the line is thus dominantly governed by the occupation effect. The transition does not broaden visibly with $T$ and its integral intensity reaches the maximum between 40 and 120~K. The observed changes in the transition intensity are likely due to the thermal redistribution of carriers between LLs. Notably, at higher temperatures the spread of the Fermi-Dirac distribution ($4k_BT$) exceeds even the CR line energy (energy separation between the $n=0$ and 1 LL). For the sake of completeness, we note that that the position and width of the fundamental CR line do not vary among explored samples, but the integral intensity does. This might be also due to the electron density of surface electrons varying among the samples investigated. 


While CR absorption~\cite{PalikRPP70} is a phenomenon regularly observed at low temperatures in semiconducting and semimetallic systems -- either bulk or 2D ones -- the observation of a well-defined CR absorption at room temperature is not common, unless extreme conditions of tens or even hundreds teslas are approached, see, e.g., ~\cite{ImanakaPBCM98,Imanakapss06,MiuraPRB97,NajdaPRB89}. In the overwhelming majority of cases, comprising also the conventional semiconductors such as silicon, germanium or GaAs~\cite{BagguleyPRSL61I,BagguleyPRSL61II,HopkinsPRB89}, 
the associated CR line broadening at room temperature exceeds in moderated magnetic fields the CR energy. In this respect, InSb is a notable exception, where CR absorption was observed well above the room temperature~\cite{PalikPR61,LiuPRB93}. In this particular system, it is the unusually low effective mass of electrons, and therefore, a large cyclotron energy which overcomes the intrinsic broadening due to electron-phonon interaction, otherwise particularly strong in polar semiconductors~\cite{YuFS96,Devreese13}. Graphene, with its particularly large energies of optical phonon modes, is another remarkable example of materials with CR absorption observed at room temperature and low magnetic fields~\cite{OrlitaPRL08II,NakamuraPRB20}. 

The thorough description of scattering mechanisms is a complex theoretical task, notably in the case of the Landau quantization, and goes beyond the scope of this paper. Nevertheless, we may compare our experimental observations with behavior of other relevant materials. With the strong optical phonon modes (Fig.~\ref{Ellipsometry} and \cite{JiangPRB20}), the BST2S TI can be classified as a polar semiconductor~\cite{Devreese13}. In such materials, one can expect that the scattering of electrons on optical phonons should dominate at high temperatures over other relevant mechanisms, such as the scattering on charge impurities or acoustic phonons~\cite{YuFS96}. This is, for instance, the case of zinc-blende semiconductors where the scattering of electrons on LO phonons -- mediated by the Fr\"olich interaction -- is known as the leading relaxation mechanism in the limit of high temperatures. In GaAs, the electron mobility (relaxation time) drops as $\mu \propto T^{-2.3}$~\cite{BlakemoreJAP82} at ambient conditions. The experimentally measured drop of mobility in GaAs is thus even faster as compared to expectations based on a simple argument about the thermally driven population of phonons ($\mu \propto 1/T$).

The absence of any visible thermal broadening of the CR resonance of electrons on the surface of BST2S TI suggests that scattering of electrons on phonons -- acoustic or optical ones -- is not the leading relaxation mechanism, even when these phonon modes are largely populated at elevated temperatures. This is consistent with other experimental works~\cite{PanPRL12,PanPRB13,SchafgansPRB12}, as well as with available numerical studies in which an unusually small Fr\"olich constant was found~\cite{HeidSR17}. This enables the existence of long-lived quasi-particles even at elevated temperatures. Having electron-phonon coupling excluded as the main source of scattering, one may speculate about the dominant relaxation mechanism responsible for the observed width of the CR mode. These might be, for instance, neutral point defects. 

To conclude, unlike bulk states, the surface states of tin-doped BiSbTe$_2$S exhibit remarkable resistance to thermally induced scattering processes. As a result, electrons in these surface states behave as long-lived quasiparticles even at room temperature. The experimental demonstration of the ability to control electronic surface states in a topological insulator by means of a moderate magnetic field at room temperature brings us closer to the long-sought development of electronic and optoelectronic devices based on topological materials, but it also opens new paths for curiosity-driven fundamental research of solids at ambient conditions.

In this context, one may imagine, for instance, the realization of magneto-statically biased non-reciprocal optical isolators. Such devices have already been experimentally demonstrated~\cite{TamagnoneNC16}, with graphene as an active medium. The BST2S TI, with well-defined surface states, could offer a more practical alternative by eliminating the need for epitaxial growth and subsequent transfer of graphene. Furthermore, the observed room-temperature Landau quantization and the exceptionally weak electron-phonon interaction in BST2S suggest the potential to realize the QHE at room temperature under high, yet experimentally attainable, magnetic fields. To date, room-temperature QHE has been convincingly observed only in high-field experiments on graphene~\cite{NovoselovScience07}, enabled by its massless Dirac-type LL spectrum -- also present in BST2S surface states -- which features a large separation between the $n = 0$ and 1 levels and thus stabilizes the robust $v = 2$ regime. In graphene, the interaction of electrons with (high-energy) optical phonons can be seen as virtually missing, but the scattering on acoustic phonons has been identified as a limiting factor in high-mobility devices~\cite{VaqueroNC23}.

\begin{acknowledgments}
We acknowledge discussions with D. M. Basko. The authors also acknowledge the support of the LNCMI-CNRS in Grenoble, a member of the European Magnetic Field Laboratory (EMFL).  The work has been supported by the ANR project TEASER (ANR-24-CE24-4830). X.S. and B.P. acknowledge support from ANR Grant No. ANR-20-CE30-0015-01. The work has been supported by the exchange programme PHC ORCHID (50852UC). R.S. acknowledges the financial support provided by the Ministry of Science and Technology
in Taiwan under Project No. NSTC (113-2124-M-001-045- MY3 and 113-2124-M-001-003), Financial support from the
Center of Atomic Initiative for New Materials (AI-Mat), National Taiwan University, (Project No. 113L900801).and
Academia Sinica for the budget of AS- iMATE-113-12.  M.P. acknowledges support from the European Union (ERC TERAPLASM No. 101053716) and the CENTERA2, FENG.02.01-IP.05- T004/23 project funded within the IRA program of the FNP Poland, co-financed by the EU FENG Programme.
\end{acknowledgments}


%

\newpage
\pagenumbering{gobble}

\begin{figure}[htp]
\includegraphics[page=1,trim = 17mm 17mm 17mm 17mm,width=1.0\textwidth,height=1.0\textheight]{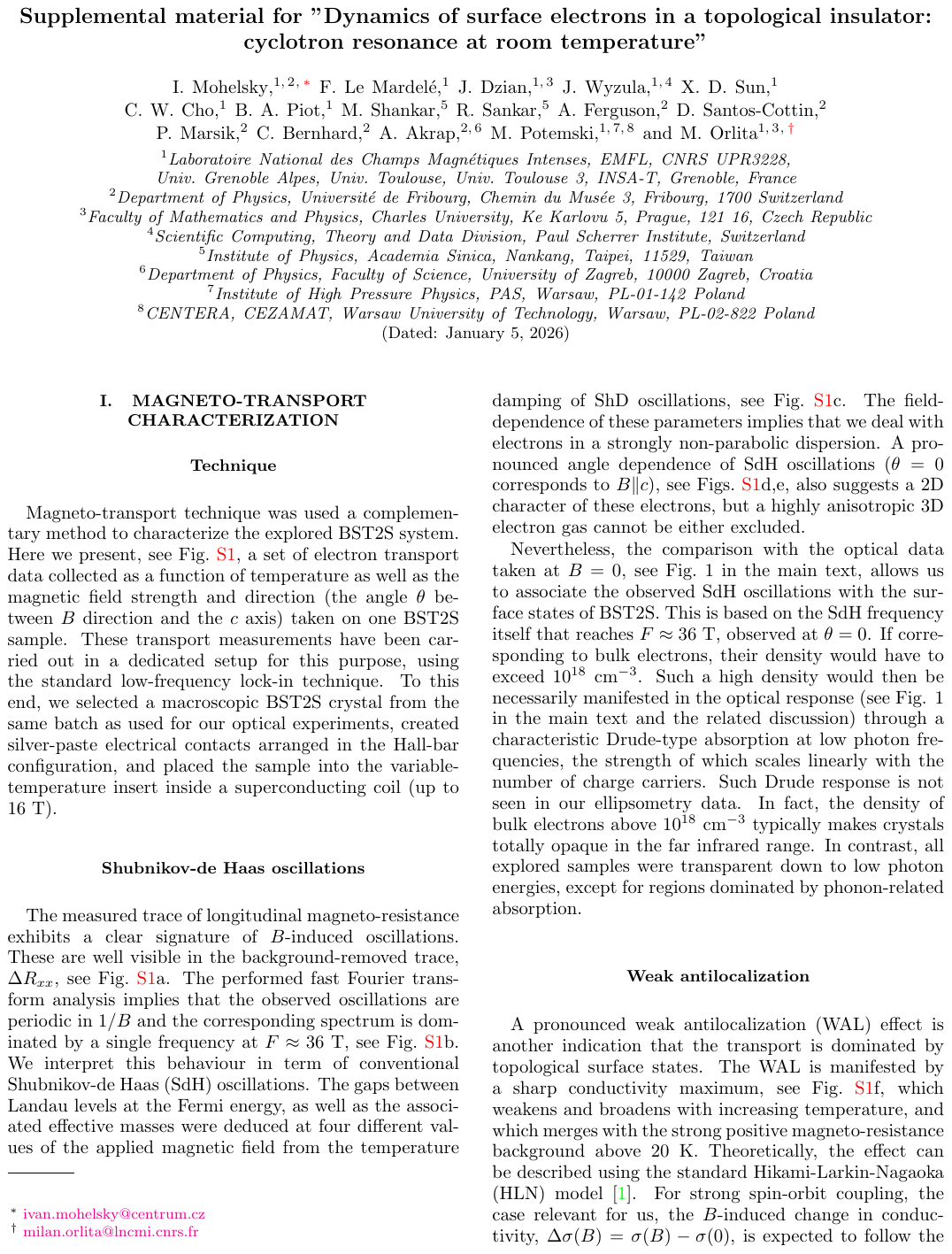}
\end{figure}

\newpage

\begin{figure}[htp]
\includegraphics[page=2,trim = 17mm 17mm 17mm 17mm, width=1.0\textwidth,height=1.0\textheight]{BiSbTe2S_PRB_SM.pdf}
\end{figure}

\newpage

\begin{figure}[htp]
\includegraphics[page=3,trim = 17mm 17mm 17mm 17mm, width=1.0\textwidth,height=1.0\textheight]{BiSbTe2S_PRB_SM.pdf}
\end{figure}

\newpage

\begin{figure}[htp]
\includegraphics[page=4,trim = 17mm 17mm 17mm 17mm, width=1.0\textwidth,height=1.0\textheight]{BiSbTe2S_PRB_SM.pdf}
\end{figure}

\end{document}